\documentclass[12pt,reqno]{amsart}

\usepackage{amssymb,amsxtra}
\usepackage{amsfonts}
\usepackage{amsmath} 
\usepackage{graphicx} 
\usepackage{layout}

\newcommand{\erre}{\mathbb{R}} 

\newcommand{\natu}{\mathbb{N}}

\newcommand{\n}{\noindent} 
\newcommand{\vs}{\vspace{0.5cm}}
\newcommand{\f}{\frac}
\newcommand{\ba}{\begin{eqnarray}} 
\newcommand{\ea}{\end{eqnarray}}
\newcommand{\be}{\begin{equation}}
 \newcommand{\ee}{\end{equation}}
\newcommand{\bdm}{\begin{displaymath}}
\newcommand{\edm}{\end{displaymath}} 
\newcommand{\brr}{\begin{array}}
\newcommand{\err}{\end{array}}

\newcommand{\bml}{\begin{gather}} \newcommand{\eml}{\end{gather}}

\newcommand{\spaz}{\vspace{.5cm} \noindent}

\numberwithin{equation}{section}

\begin{document}

\title[ ]{Classical behavior in quantum systems: \\ \vspace{0.1cm} the case of straight tracks in a cloud chamber\\
 \vspace*{0.6cm}}


\author{Alessandro Teta} \address{ A. Teta: Dipartimento di Matematica Pura ed
Applicata,  Universit\`a di L'Aquila}

\curraddr{Via Vetoio (Coppito 1), 67010 L'Aquila, Italy}

\email{teta@univaq.it}

{\maketitle}

\vs




\begin{abstract}
The aim of this review is to discuss in a pedagogical way the problem of the emergence  of a classical behavior  in certain physical systems which, in principle, are correctly described by quantum mechanics. It is stressed that the  limit  $\hbar \rightarrow 0$ is not sufficient and the crucial role played by the environment must be taken into account. 
In particular it is recalled the old problem raised by Mott in 1929 (\cite{m}) concerning  the straight tracks observed in a cloud chamber, produced by an $\alpha$-particle emitted by a source in the form of a spherical wave. The conceptual  relevance of the problem for a clearer understanding   of the classical limit is discussed in a historical  perspective. Moreover  a simple mathematical model is proposed, where the result of Mott is obtained in a rigorous mathematical way.

\end{abstract}

\vs
\vs

\section{Introduction }

\vs
\n
Quantum mechanics (QM) is a well established theory formulated in the first half of the last century to account for the behavior of microscopic systems, i.e. systems in which a typical action is comparable in size with the Planck's constant $\hbar$. Nevertheless in the non relativistic regime QM is a universal theory  and it can also be applied to the macroscopic world or in general to situations where the classical description is valid to a high degree of accuracy. 
In such cases, a relevant question is to guarantee that the "correspondence principle" holds, i.e.   to show that the quantum mechanical description reduces to the classical one through a suitable limit procedure. It is well known that,  under appropriate conditions on the initial state of the system, this reduction can be obtained taking, roughly speaking,  the limit $\hbar \rightarrow 0$. 
As a matter of fact, in some relevant situations such procedure turns out to be inadequate and the role of the environment  
must be taken into account in order to show how  a classical behavior can appear in  a quantum system. The corresponding   dynamical mechanism is known in the physical literature under the name of decoherence (see e.g. \cite{gjkksz}). In particular we shall focus on the specific case considered by Mott (\cite{m}) concerning a possible explanation of the straight tracks observed in a cloud chamber.

\n
In this review the analysis of the problem will be carried out in a historical and pedagogical perspective and some emphasis will be given to the conceptual implications of the different proposed approaches.

\n
We shall start in section 2 recalling the basic rules of QM with no pretense of full rigor and generality. The only aim is to underline the crucial points of the so-called standard formulation (see e.g. \cite{s}), with a brief discussion about the difficulty arising from the measurement problem and in general from the existence of superposition states for macroscopic objects. 
Because of lack of space, we will not mention  other possible formulations of the theory, like Bohmian mechanics (\cite{dgz}),  GRW theory (\cite{grw}) or  many worlds formulation (\cite{e}). For the reader interested in the conceptual foundation of QM we refer to the books \cite{b}, \cite{g}.

\n
In section 3 we  discuss the problem of the classical limit and the important role of decoherence induced by the environment which results in the dynamical suppression of the quantum correlations present in a  superposition state.

\n
In section 4 we shall briefly describe how a cloud chamber works and analyse the problem of the appearence of  straight tracks produced by the passage of an $\alpha$-particle. In particular we shall compare the explanations based on the wave packet collapse from one side and on the quantum mechanical analysis of the $\alpha$-particle plus the atoms of the chamber on the other side. The latter approach is the one proposed by Mott.

\n
In section 5, following the same line of thought, we consider a simplified one dimensional model of a  cloud chamber  and  we describe  a theorem, proved in (\cite{dft}), which can be considered as a mathematical proof, in a simple case, of the heuristic arguments given by Mott. Some final remarks will conclude the paper.

\vs
\section{Basic rules of QM}

\vs
\n
As it is well known, Classical Mechanics (CM) provides a description of the motion of a system of $n$ point particles in $\erre^d$, $d=1,2,3$,   based  at kinematical level  on their positions $x_1, \ldots , x_n$ and momenta $p_1,\ldots,p_n$. 
This means that at each time the state of the system is identified by a point in a $2 d n$-dimensional space, called  phase space.
 Given the initial state and the forces acting on the system, in principle it is possible to solve the Newton's equation and determine the state  of the system  at any other time. Furthermore 
a classical observable is represented by a smooth real function defined on the phase space of the system.   In particular the predicted value of an  observable $f$ when the system is in the state $(x_1, \ldots , x_n,p_1,\ldots,p_n)$ is simply given by $f(x_1, \ldots , x_n,p_1,\ldots,p_n)$. An important point is that in CM the algebra of observables is commutative and this fact  allows a prediction  of the value of all the observables relative to a given system.

\n
At the beginning of the twentieth century it became clear that CM, and classical electromagnetism, fail when applied to atomic systems and it was realized that a radically new description was required. The new theory was elaborated in the years 1925-26 due to the work of Heisenberg and Schr\"{o}dinger and  it was formulated in a rigorous mathematical language by von Neumann in 1932. 
The crucial point is that QM is based on a new and more abstract kinematics. 
More precisely the commutative algebra of observables typical of CM must be replaced by a non commutative algebra of selfadjoint operators in a suitable Hilbert space and, at each time,   
the state of the system is given by  a vector in the same space. 
We summarize here the basic rules for a system of $n$ quantum particles in $\erre^d$, $d=1,2,3$, avoiding  generality and technical difficulties  and in particular neglecting the specific rules required for systems of identical particles.

\vs

1.  All the information on the system is encoded in the wave function $\psi_t (x_1, \ldots ,x_n)$, $x_j \in \erre^d$,  which is an element of the Hilbert space $L^2(\erre^{dn})$ with $\|\psi_t\|=1$. The wave function $\psi_t$ is referred to as the (pure) state of the system at time $t$.

2. Given the initial state $\psi_0$, the state at time $t$ is the solution of the Schr\"{o}dinger equation 
\be
i \hbar \f{\partial \psi_t}{\partial t}= \sum_{j=1}^{n} \f{- \hbar^2}{2m_j} \Delta_j \psi_t + V(x_1, \ldots,x_n) \psi_t
\ee
with initial datum $\psi_0$, where $m_j$ is the mass of the $j$-th particle, $\Delta_j$ denotes the Laplace operator relative to the coordinates of the $j$-th particle and $V$ is the interaction potential.

3.  An observable relative to the system is represented by a selfadjoint operator in $L^2(\erre^{dn})$.   In a system made of a single quantum particle, simple examples of quantum observables are  position and  momentum.  The position  is represented by $\hat{x}_{k}$, defined as the multiplication operator by $x_{k}$, $k=1,..,d$, where $x_{k}$ denotes the $k$-th component of the position of the particle.  Analogously  the momentum  is represented by the differential operator $\hat{p}_{k}=-i\hbar \f{\partial}{\partial x_{k}}$, $k=1,..,d$. One can easily check that the two observables do not commute and in fact, at least  formally, they satisfy the Heisenberg commutation relations $[\hat{x}_{k}, \hat{p}_l ] = i \hbar \delta_{k l} I$, where $I$ denotes the identity operator. 

4. The predictions of the theory are given by  Born's rule and, in general, are of probabilistic type. 
In the special case of the position observable relative to a system made of one quantum particle  the Born's rule reduces to
\be\label{br}
{\mathcal P}(x \in \Omega; \; \psi)= \int_{\Omega} \!dx \, |\psi (x)|^2
\ee
where the l.h.s. denotes the probability that the position of the  particle described by the state $\psi$ is found in a  set $\Omega \subset \erre^d$.  
The prescription can be easily extended to the case of  observables different from the position making use of the spectral theorem for selfadjoint operators.

\vs

\n
We list here some comments on the above rules.

i)  As already mentioned, QM is formulated as a universal theory (in the non relativistic regime) and in principle it can be used to describe both micro and macro systems.

ii)  The predictions of the theory are in excellent agreement with experiments.

iii) Except in some special cases, the Born's rule gives only probabilistic predictions. In particular formula (\ref{br}) means that the theory can only predict the statistical distribution of the detected positions in a large number of experiments made in identical conditions. We notice  that a quantum particle, when detected in a single experiment, always appears localized in a well defined position which, in general, cannot be predicted by the theory.

\n
Since we have assumed that the state $\psi$ gives a complete information on the system, the above probability cannot be considered epistemic, as it is common in classical physics.  Such  completeness assumption is typical of the standard formulation of QM.

iv)  For a single observable $A$ it is always possible to find a state $\psi$ such that $\Delta_{\psi} A$ is arbitrary small, where $\Delta_{\psi} A$ denotes the mean square deviation of the statistical distribution of the possible values of $A$ in the state $\psi$. This means that the values of the observable $A$ in the state $\psi $ can be predicted with arbitrary accuracy.  On the other hand, the non commutative character of the algebra of observables implies that one cannot predict, with arbitrary accuracy, the value of all the observables relative to a given system. In particular, for a quantum particle in the state $\psi(x)$  the Heisenberg uncertainty principle states
\be\label{ind}
\Delta_{\psi} \hat{x}_k \, \Delta_{\psi} \hat{p}_l \geq \f{\hbar}{2} \delta_{kl}
\ee

v)  A crucial point of the theory is the linearity. This means that if $\psi_1 (x)$ and $\psi_2(x)$ are two  states then also the  sum $\psi_1(x)+\psi_2(x)$, suitably normalized, is a possible state (superposition principle). This apparently trivial fact has strong physical consequences, due to the fact that the predictions (see (\ref{br})) are given  by a quadratic expression with respect to the state. 
In particular the probability density for the position is 
\be\label{sp}
|\psi_1(x)+\psi_2(x)|^2 = |\psi_1(x)|^2 + |\psi_2 (x)|^2 + 2 \Re  \left(\psi_1 (x) \overline{\psi_2 (x)} \right)
\ee
From formula (\ref{sp}) it is clear that the situation described by $\psi_1(x)+\psi_2(x)$ cannot be considered in any sense as the "sum" of the situations described by $\psi_1 (x)$ and $\psi_2(x)$ separately. In particular the last term in (\ref{sp}) is responsible for the appearance of interference effects, typical of waves, in the statistical distribution of the detected positions in a large number of identical experiments. Such effects can be directly observed in the so called two-slit experiment. 

\n
We notice  that a completely different situation occurs when we know that the system is in the state $\psi_1(x)$ with probability $p_1$ and in the state $\psi_2(x)$ with probability $p_2$, where $p_1 + p_2=1$. In such case the probability involved has an epistemic nature and the system is described by a so called classical statistical mixture of the two pure states $\psi_1(x)$ and $\psi_2(x)$.

vi)  Another important issue  is the occurrence of "entanglement" for systems composed by more than one particle.  An entangled state is a state that cannot be factorized in a product of one-particle states. 
At a kinematical level, this means that if  a system is  described by an entangled state $\psi(x_1,\ldots,x_n)$, it is not possible to associate a definite (pure) state to each subsystem. The situation is again radically different from the classical case and it is the origin of the "non local effects" which can be produced on a subsystem $A$ acting on another spatially separated subsystem $B$.

vii)  It should be stressed that superposition principle and entanglement are the most peculiar aspects of the quantum description. They are responsible for the strange behavior of the quantum objects but also for the extraordinary success of the theory in a large variety of applications.

viii)  It is  clear from the above remarks that the behavior of a quantum object  cannot  be reduced to that of a classical particle or a classical wave.  In the early days of QM this fact was interpreted in terms of the so-called wave-particle duality, i.e. it was claimed that  a quantum object behaves like a   wave or like a particle depending on the given physical situation and the two aspects are mutually exclusive. This duality was considered a special case of a  general philosophical view known as Bohr's complementarity principle. At that time such view was a reasonable attempt to understand  the new  phenomenology in terms of a more traditional intuition based on concepts of classical physics.   After more than eighty years from the birth of QM  it seems reasonable to accept that the behavior of  a quantum object is something completely different from the behavior of a classical object. At a psychological  level,   it would  probably be  helpful   to use a new name, e.g.  "quanton", as discussed in \cite{bl}.

\vs
\n
Even in a short introduction to the standard formulation of QM, it is worth mentioning some conceptual difficulties arising in the description of the measurement process.

\n 
In order to exemplify the problem, let us consider the simple case of a quantum particle described by the state $\psi_j$, $j=1,2$, where $\psi_j$ is a  wave packet well localized around the position $x_j$. The quantum particle is supposed to interact with a macroscopic measurement apparatus designed to measure the position of the particle. 
Before the interaction the apparatus is described by the state $\Phi_0$, corresponding to the "index" at rest.
 
\n
Let us assume that after the interaction the system is still in $\psi_j$ and the apparatus is in $\Phi_j$, i.e. a state corresponding to the index in the position $j$ and showing that the particle position is $x_j$. 
In other words, the situation after the measurement  for the whole system + apparatus is described by the product state $\psi_j \otimes \Phi_j$. It is intuitively clear that this is the  typical response of an apparatus and   we remark that model interactions between system and apparatus producing situations of this type can be effectively constructed. The problem arises if the system is initially in the superposition state $\psi_1 + \psi_2$. As a pure consequence of the linearity of the evolution, after the interaction the state of system + apparatus is $\psi_1 \otimes \Phi_1 + \psi_2 \otimes \Phi_2$, which is a typical entangled state corresponding to the superposition of $\psi_1 \otimes \Phi_1$ and  $ \psi_2 \otimes \Phi_2$. In principle this means that one cannot assign a definite position to the index of the apparatus and some interference effects could be observed for the position of the index, like in the case of the two-slit experiment.  Obviously the situation is unsatisfactory since we always see the index of a macroscopic  apparatus in a well defined position.

\n
In order to avoid the above difficulty one possibility is to invoke  an additional rule called wave packet reduction. Roughly speaking the rule simply states that the macroscopic apparatus must behave classically and it cannot be found in a superposition state. From a  technical point of view this means that after the measurement the  superposition state $\psi_1 \otimes \Phi_1 + \psi_2 \otimes \Phi_2$ instantaneously reduces to   the classical statistical mixture of the two possible states $\psi_1 \otimes \Phi_1 $ and $ \psi_2 \otimes \Phi_2$. The critical aspect is that such reduction cannot be described by a dynamics governed by the Schr\"{o}dinger equation and  one  has to admit that in some situations the Schr\"{o}dinger dynamics fails and must be replaced by a different dynamical rule. 

\n
This is a rather strange fact  since, after all,  a macroscopic apparatus is made of atoms whose behavior is described  by the Schr\"{o}dinger equation. Moreover it is not precisely prescribed     when and under which circumstances  the wave packet reduction should occur and this introduces  a further ambiguity in the application of the rule.

\n
Another possible point of view is to insist that the only existing dynamical rule is the Schr\"{o}dinger dynamics. In this case one has to consider specific models of system + apparatus described by the Schr\"{o}dinger equation. Then the problem becomes to prove that the replacement of the superposition state with the statistical mixture can be done at the cost of an error so small that it would be practically undetectable in a real experiment. Such kind of models have been effectively constructed and some interesting results in this direction are available (\cite{he}, \cite{se}).

\n
This suggests that in practical circumstances the wave packet reduction is an innocuous procedure and  the standard formulation of QM can be considered valid {\em for all practical purposes} (\cite{b}). 

\n
Nevertheless also with this approach  a delicate conceptual difficulty remains.  Indeed QM would be a fundamental theory which acquires a precise meaning only through an approximation procedure, although very accurate, depending on the specific model employed.

\n
In conclusion one has to admit that the description of the measurement process within the standard formulation of QM is not completely satisfactory 
from a conceptual point of view and it is not surprising that the question has stimulated an   intense epistemological debate  which is still active.

\vs
\vs
\section{The classical limit}

\vs
\n
There are many examples in  the history of physics of a theory $T_1$, describing very  accurately a certain class of phenomena, which turns out inadequate when applied to an emerging new set of experimental data. In these cases one tries to construct a new theory $T_2$ for the explanation of the experimental data with the  requirement  that $T_2$ must be "more general" that $T_1$, i.e. it must  reduce to $T_1$  in a suitable limit. The underlying idea is that  a physical theory is not only a set of rules useful to predict the results of experiments but it is also supposed to provide a  unified description of a wider and wider portion  of the physical reality. 
A typical example is the special relativity as a Lorentz-invariant  generalization of CM. In this case it is easy to see that the former reduces to the latter in the limit $" c \rightarrow \infty"$, where $c$ is the speed of light. From the technical and conceptual point of view the limit procedure is greatly simplified  by the fact that both theories deal with  point particles  characterized by position and momentum and evolving in time in ordinary space. 

\n
Another example is  wave optics as a generalization of  geometrical optics, required to account for interference and diffraction effects of light. In this case  the two theories are based on a different space-time description of  light propagation, i.e. as a wave and as a ray respectively. The technical difficulty here is to specify in which sense the propagation of a wave  can reproduce the propagation of a ray. This can be done with a natural choice of the initial datum and  exploiting the "short wave-length"  limit  for the corresponding solution of  Maxwell  equations. After a  non trivial  mathematical analysis of the problem one can conclude that in the limit   the laws of geometrical optics are recovered.

\n
The problem of the classical limit of QM is surely more delicate than the previous examples. 

\n
Apparently the situation is similar to the case of optics, since we compare a theory based on point particles propagating along  trajectories (CM) with a theory based on a kind of "wave" solution of the Schr\"{o}dinger equation (QM). Nevertheless one realizes that the situation is deeply different  if one observes that the wave representing the quantum state does not describe a physical object distributed in ordinary space like in optics. Rather it is only a probability amplitude for the position distribution of the quantum object which, when detected, is well localized in a given point. Moreover the 
superposition principle introduces a crucial  difficulty since we cannot give any definite meaning to the position of the quantum object in a superposition state.  In other words the standard formulation of QM does not provide a  space-time description of the behavior of a quantum object easily  comparable with the classical one. For these reasons the classical limit of QM is a very delicate problem both from the technical and  the conceptual point of view (see e.g. \cite{ck} for a historical analysis of the problem).

\n
The traditional approach is essentially based on the analysis of the solutions of the Schr\"{o}dinger equation in the limit  $"\hbar \rightarrow 0"$ for  a suitable choice of the initial state.   We recall that in this context the limit $"\hbar \rightarrow 0"$ simply means that the typical action of the system is negligible with respect to the Planck's constant.

\n
Usually one considers two possible initial states, chosen in analogy  with the case of optics. The first one  is the WKB state defined by an   amplitude independent of $\hbar$  and a highly oscillating phase for $\hbar$ small. In this case one can show, for $\hbar $ small and for short times, that the corresponding solution of the Schr\"{o}dinger equation has the same form, with amplitude and phase governed by the classical transport and Hamilton-Jacobi equations respectively. This means that in the limit the quantum state propagates like a classical fluid and in this sense the classical description is recovered. 

\n 
A second reasonable choice of initial state is the coherent state, i.e.  a 
wave packet well concentrated in position and momentum around a point $(x_0,p_0)$ in the classical phase space for $\hbar $ small.  One can prove that the time evolution, for $\hbar $ small and a time interval not too long, is again described by a wave packet well concentrated in position and momentum around the classical trajectory starting from $(x_0,p_0)$.

\n
A precise statement and a  mathematical proof of the above qualitative pictures have required much technical work and  many refined and detailed results are now available in the literature (\cite{r}).

\n
Despite their mathematical elegance, such kind of results cannot be considered  satisfactory for a complete understanding of the classical limit of QM. The reason is that the approach is crucially based on the choice of specific, essentially classical initial states while, in many cases, the classical behavior should emerge also starting from a genuine quantum state, like a superposition state. In these case the usual procedure $\hbar \rightarrow 0$ is not sufficient.   The difficulty was clearly recognized by Einstein in his correspondence with Born, in the years 1953-54 (\cite{bo}). The answer given by the founders of the standard formulation of QM (see e.g. the Born-Pauli correspondence in \cite{bo}) was based on the rule of the wave packet reduction and therefore it was outside the context of an evolution governed by  the Schr\"{o}dinger equation. In any case at that time the question was  considered  relevant only at foundational level  without  practical consequences. Starting around 1980 some remarkable experimental progresses have made possible a detailed  exploration of the classical/quantum border. In particular genuine superposition states of quantum systems have been prepared and their evolution analysed in real  experiments, often in connection with problems arising in  the 
new emerging field of quantum information. At a theoretical level a new impulse was given to the construction of  models in which one can   analyse if and how much a superposition state can survive or eventually  can be reduced to a classical statistical mixture. It is important to remark that here one is interested in  a quantitative analysis of the problem. Therefore  any explanation based on the  wave packet reduction cannot help and  one is forced to consider the problem entirely in the context of the Schr\"{o}dinger equation.

\n
The idea behind this theoretical analysis is that the quantum coherence  present in a genuine superposition state is very fragile. This implies that even a weak interaction of the system with the environment can destroy the superposition state and a classical behavior of the system can emerge. The dynamical mechanism producing the reduction or the suppression of the quantum coherence for a system interacting with the environment is usually called decoherence. In the last years a considerable work has been done to control and quantify the decoherence induced by the environment in many interesting situations. Here we want to  mention the analysis of the reduction of the interference effects for a heavy particle due to the scattering of light particles, in a typical  two-slit experiment (\cite{jz}, \cite{hs}). Another  interesting case is the explanation of the molecular localization for  pyramidal molecules, like $NH_3$, due to the dipole-dipole interaction among the molecules (\cite{cj}, \cite{jpt}, \cite{gms}). 
In the next section we shall concentrate on a further example, i.e the emergence of classical trajectories in a cloud chamber, which seems particularly interesting for a better understanding of the classical limit of QM.

\vs

\vs
\section{$\alpha$-particle in a cloud chamber}

\vs
\n
A cloud chamber is an experimental apparatus which was constructed and improved by Wilson in the years 1911-12 (see e.g. \cite{lr} for a description of the original apparatus). The cloud chamber proved to be particularly useful for the investigation of the properties of  various atomic and sub-atomic particles. 
In particular it was used to observe the trajectory of an $\alpha$-particle emitted by a radioactive source of radium placed in the chamber. The way of working  of the apparatus can be schematically described as follows. The chamber is filled by a supersaturated  vapour which can undergo local phase transitions induced by the exchange of even  a small amount  of  energy. The $\alpha$-particle released  by the source propagates   and interacts with the atoms of the gas in the chamber inducing ionization. The ionized atoms  produce a local phase transition of the gas  and therefore the formation of small drops of cloud near their positions. The sequence of such drops is the visible track that one can directly observe in the chamber. The tracks have usually the form of straight lines (or of curved lines whenever  magnetic and/or electric  fields are  applied) and they can be considered as the experimental manifestation of the "trajectory" of the $\alpha$-particle. 

\n
The theoretical explanation of this phenomenology must be given in the framework of QM due to the microscopic character of the processes involved. In particular the successful analysis of the radioactive decay of a nucleus  with emission of an $\alpha$-particle was achieved  by Gamow (\cite{ga}) and by Condon and Gurney (\cite{cg}). It was shown that the emitted $\alpha$-particle is described by a wave function having the form of a spherical wave, centered in the source,  and isotropically propagating in space.

\n
It was soon realized that this fact is apparently in contrast with the experimentally observed straight tracks in the cloud chamber, which are typical of a particle behavior.  The problem was intensively discussed in the early days of QM since it was a serious and interesting test for the new theory.  In particular it was considered important  to clarify the connection between the particle-like and the wave-like behavior of a quantum system.

\n
The problem was summarized by  Born (\cite{ep}) in his communication at the Solvay Conference in 1927: "If a spherical wave is associated with every act of emission, how can  it be understood that the trace of the $\alpha$-particle appears as an (almost) straight line?"

\n
 A further clear enunciation of the problem was given by Darwin (\cite{d}): "Consider for example one of the most striking manifestations of particle characters, the ray tracks of $\alpha$-particles in a Wilson cloud chamber. We have to connect this in some way with the theory of radioactive disintegration as presented by Gamow. In that theory the radium nucleus contains what we  may call an $\alpha$-wave which is slowly and continuously leaking out as a spherical wave."

\n
We finally mention the words of  Mott (\cite{m}): "It is a little difficult to picture how it is that an outgoing spherical wave can produce a straight line; we think intuitively that it should ionize atoms at random throughout space".

\n
It was clearly recognized by Heisenberg in his Chicago lectures in 1929 (\cite{h})  that in the framework of the standard formulation of QM one has two possible points of view to reconcile  theory with experimental data.  The first one is based on the wave packet collapse and the second on the consideration of the Schr\"{o}dinger equation for the  $\alpha$-particle and 
the atoms of the gas. 

\n
According to the first point of view the gas in the chamber must be considered as a (classical) measurement apparatus which repeatedly  measures the position of the $\alpha$-particle. The result of the   measurement operated by the first atom of the gas is the reduction of the original spherical wave to a wave packet localized near the position of the atom and propagating along the line joining the source with the atom. Such wave packet is subject to an unavoidable spreading which is again  reduced by the position measurement of the next atom and so on. This mechanism would explain the appearance of an almost classical trajectory of the $\alpha$-particle. The explanation is surely simple and effective  but on the other hand it can be criticized. In fact, even if one accepts the rule of the collapse, it is not clear why a microscopic system, like an atom of the gas, should behave as a measurement apparatus. At least it should be  specified under which physical conditions this can happen.

\n
If one takes seriously  the second point of view the atoms of the gas are considered as quantum objects together with the $\alpha$-particle and one has to derive   the effect, i.e. the straight tracks, from the Schr\"{o}dinger equation applied to the whole quantum system. In this context a crucial role is played by the entanglement produced by the interaction between the $\alpha$-particle and each atom of the gas. The basic idea was  expressed in \cite{d}: "Before the very first collision (the wave function) can be represented as the product of a spherical wave for the $\alpha$-particle, by a set of more or less stationary waves for the atoms. But the first collision changes this product into a function in which the two types of coordinate are inextricably mixed, and every subsequent collision makes it worse."  Such complicated function contains a phase factor and "without in the least seeing the details, it looks quite natural to expect that this phase factor will have some special character, such as vanishing, when the various co-ordinates  satisfy a condition of collinearity. So without pretending to have mastered the details, we can understand how it is possible that the $\psi$ function, so to speak, not to know in what direction  the track is to be, but yet to insist that it should be a straight line."

\n
The program enunciated by Darwin was concretely carried out by Mott in \cite{m}. He studied a simplified model with only two atoms of the gas placed at fixed positions $a_1$, $a_2$  and the $\alpha$-particle initially described by a spherical wave centered in the point $O$, with $|Oa_2| >|Oa_1|$. He computed the ionization probability of both atoms  and he found that such probability is essentially zero unless the second atom in $a_2$ lies on the straight line joining $O$ with $a_1$.  The basic idea is the following.  The ionization of the first atom in $a_1$ produces a well localized wave packet starting from $a_1$ with momentum directed from $O$ to $a_1$. Therefore if we require that also the second atom is ionized the only possibility is that its position is on the trajectory of the wave packet.  
The analysis exploited a perturbative expansion up to second order of the solution of  the time-independent Schr\"{o}dinger equation and made use of the stationary phase method, essentially along the line suggested in \cite{d}. 
Although the result was obtained only through heuristic arguments,  it was based  on a beautiful physical intuition and the approach to the problem must be surely considered pioneering.  As a matter of fact the work was rather neglected for long time.  This was  probably a consequence of a too rigid acceptance of the standard formulation of QM, rather common in the scientific community during the first years of the theory.  In  particular a special emphasis   was given to the role of the act of measurement, and then to the rule of the wave packet collapse, in order to solve possible conceptual difficulties of the theory.  In this context it is clear that the approach proposed by Mott, entirely based on the study of the Schr\"{o}dinger equation,  could appear rather misleading.

\n
In particular it could obscure the role of the act of observation which should be necessarily invoked at a certain point. We notice that in principle the objection is correct but, nevertheless, it is not irrelevant that this point can be postponed. In this sense  Mott's approach is  more interesting from the physical point of view since it allows a quantitative analysis of the processes involved which is excluded with the wave packet collapse.

\vs
\vs
\section{A one dimensional model}

\vs
\n
In recent years further elaborations on  the subject of the straight tracks in a cloud chamber  have been worked out (\cite{p}, \cite{br}, \cite{ccf}, \cite{cl}, \cite{ha}). Here we want to describe a simplified one dimensional model consisting of a test particle and two harmonic oscillators, where a rigorous mathematical analysis can be carried out (\cite{dft}). A superposition state of two wave packets centered in the origin with opposite momentum plays the role of the spherical wave of the $\alpha$-particle and the oscillators replace the atoms to be ionized.  Under suitable assumptions on the physical parameters of the model, a detailed analysis of the time evolution of the system can be performed using time dependent perturbation theory. The result is  a quantitative  estimate of the joint excitation probability of the oscillators. In particular it is shown   that such probability is essentially zero if the two oscillators 
are placed on  opposite sides of the origin, while it has a finite, non-zero value in the other case.  
The work is entirely within the line of reasoning of Mott and it develops  some aspects of the model which are not considered in \cite{m}.  In particular the analysis is performed following the time evolution of the system and, moreover, quantitative estimates are given showing precisely under which physical conditions the effect can be seen. 

\n
Let us introduce the model. We denote by $\Psi(R,r_1,r_2,t)$ the wave function of the system, where $R$ is the coordinate of the test particle and $r_1, r_2$ the coordinates of the oscillators placed in positions $a_1, a_2$ respectively, with $|a_2|>a_1>0$. The Schr\"{o}dinger equation reads

\ba
&&i \hbar \f{\partial \Psi}{\partial t} = H \Psi \label{eq}\\
&&H= -\f{\hbar^2}{2M} \Delta_R \! -\! \f{\hbar^2}{2m} \Delta_{r_1} \! +\!\f{1}{2}m \omega^2 (r_1 \!-\! a_1)^2 \!-\! \f{\hbar^2}{2m} \Delta_{r_2} \!+\!\f{1}{2}m \omega^2 (r_2\! -\! a_2)^2 \label{H_0}\nonumber\\
&&+ \lambda V(\delta^{-1}(R-r_1)) + \lambda V(\delta^{-1}(R-r_2)) \label{H_1}
\ea
where $M$ is the mass of the test particle, $m, \omega$ are the mass and the frequency of the oscillators, $V$ is an interaction potential and $\lambda , \delta $ two positive parameters. As initial state  we choose 
\ba
&&  \Psi_0(R,r_1,r_2)= \psi(R) \phi^{a_1}_{0} (r_1) \phi^{a_2}_{0}(r_2) \label{statoin}
\ea
where  $\psi(R)$ is the one dimensional spherical wave for the test  particle
\ba
&&\psi(R)= \psi^{+}(R) + \psi^{-}(R)  \equiv    \f{\mathcal{N}}{\sqrt{\sigma}} e^{- \f{R^2}{2 \sigma^2}} e^{+ i \f{P_0}{\hbar} R}    +    \f{\mathcal{N}}{\sqrt{\sigma}} e^{- \f{R^2}{2 \sigma^2}} e^{- i \f{P_0}{\hbar} R} \label{p}
\ea
with $\sigma, P_0 >0$ and $\mathcal{N}$ normalization factor. Moreover $\phi_{n_i}^{a_i} (r_i)$, $i=1,2$, denotes the eigenfunction of the oscillator in $a_i$ corresponding to the energy level indexed by $n_i \in \natu$. 

\n
We remark that in absence of  interaction  the test particle would be described by  the free evolution of the superposition state (\ref{p}), i.e. a coherent superposition of two wave packets, one propagating on the right and the other on the left.  Therefore, according to the standard formulation of QM,  the test particle  would be completely delocalized and its position would have no meaning. 

\n  
 The relevant object of the analysis is $\mathcal{P}_{n_1 n_2} (t)$, i.e. the probability that both oscillators are excited in the states labeled by $n_1, n_2 \neq 0$ at time $t$. From a direct application of the Born's rule one has 
\be
 \mathcal{P}_{\!n_1 n_2 \!}(t)= \int_{\erre} \!\! dR \;  |f_{n_1 n_2} (R,t) |^2
 \ee
where $f_{n_1 n_2} (R,t)$ are the coefficients of the expansion
\be
\Psi (R,r_1,r_2,t) = \sum_{n_1,n_2} f_{n_1 n_2}(R,t) \phi^{a_1}_{n_1}(r_1) \phi^{a_2}_{n_2}(r_2)
\ee

\n 
The computation is done for $t>\f{|a_2|}{v_0}\equiv \tau_2$,  where $v_0=\f{P_0}{M}$, and $\tau_2$ is the time required to a classical particle to go  from the origin to $a_2$. An important point is to specify the set of assumptions on the physical parameters of the model. In particular we assume

\n
(1) $\lambda_0 \equiv \f{\lambda}{Mv_0^2} \ll 1$.

\n
(2) The dimensionless quantities $\f{m}{M}$, $\f{\hbar \omega}{Mv_0^2}$, $\f{\sigma}{|a_i|}$, $\f{\delta}{|a_i|}$, $\f{v_0}{\omega |a_i|}$, $i=1,2$, are all $O(\varepsilon)$, where $\lambda_0 \ll \varepsilon \ll 1$.

\n
For a discussion of the physical meaning of (1),  (2) we refer to \cite{dft}. The mathematical result can be formulated as follows. 

\vs
\n
 {\bf Theorem 1.} 
Let us assume (1),(2) and $n_1, n_2 \neq 0$. Then, up to second order in perturbation theory and for $t>\tau_2$,  one has

\n
i) for $a_2 < 0<a_1$ 
\be
\mathcal{P}_{n_1 n_2}(t) \leq \varepsilon^k \left(\! \f{\lambda_0}{\varepsilon} \! \right)^4  A^{(k)}_{n_1 n_2} (t), \;\;\;\;\;\mbox{for any} \;\;\;k \in \natu
\ee
ii) for $0<a_1 <a_2$
\be
\mathcal{P}_{n_1 n_2}(t)= \left(\! \f{\lambda_0}{\varepsilon} \! \right)^4 B_{n_1 n_2}(t)
\ee
where $A^{(k)}_{n_1 n_2} (t)$, $B_{n_1 n_2}(t)$, for $t$ of order of $\tau_2$, are quantities of order one which are explicitly estimated. 

 \vs
 \n
The theorem clearly expresses the fact that  the probability $\mathcal{P}_{n_1 n_2}(t)$ computed in case i) is negligible with respect to the same probability computed in case ii). For the proof we refer to \cite{dft}  and here we only add some comments on the meaning of the result.

\n
In general QM can only predict  the probabilities of the possible events during the evolution of  a given system. In our model in case i) we are concerned with four possibilities, which we may call possible histories:  no oscillators is excited, only the oscillator on the right is excited, only the oscillator on the left is excited, both oscillators are excited. The theorem states that the last history has a negligible probability and then we can only observe just one excited oscillator. If we observe that the oscillator on the right (resp. on the left) is excited we can affirm that the test particle is localized on the right (resp. on the left). This means that  the interaction determines the localization of the test particle (on the right or on the left) while, before the interaction, the test particle was completely delocalized and no meaning could be given to its position. In this rather indirect sense the test particle behaves like a classical particle propagating along a well defined trajectory.

\n
We emphasize that the result strongly depends on the assumptions (1),(2) and one should expect a completely different behavior if other physical assumptions are introduced.  
This fact once again shows the difference with respect to the approach based on the wave packet collapse.

\vs
\vs
\section{Concluding remarks} 

\vs
\n
In the previous sections 4, 5 we have discussed a specific case of classical behavior emerging in a quantum system. We have emphasized  that the pioneering approach introduced by  Mott was  rather neglected  for a long period and only  recently  rediscovered and applied to a number of other situations, where the environment plays a role in the suppression of quantum coherence (decoherence theory). 
We have also underlined that the birth and development of such interesting field of research was prevented for long time on the basis of pure ideological motivations, corresponding to a too rigid interpretation of the theory. 
In any case we have now a well established set of results in the physical literature which allows a  quantitative description of the emergence of classical behavior in many concrete situations.

\n
It is worth mentioning that such description  is surely valid {\em for all practical purposes}, in complete analogy with 
the problem of the measurement process mentioned in section 2. Therefore once again one can ask whether it is satisfactory also from the conceptual point of view.

\n
The answer  depends on the idea of scientific explanation or scientific theory one has in mind. There are surely many different opinions on this question and we want just to recall  two possible views 
which, roughly speaking, can be summarized  as follows.

\n
According to a strumentalist  view, a scientific theory should not pretend to describe the objective world. Rather, it is only a set of rules, formulated in mathematical language, useful to organize the empirical experience and to predict the results of experiments. 
Such view influenced the founders (Bohr, Heisenberg, Pauli) of the standard formulation of QM and it is rather common in the physics community.

\n
On the other side there is a more realistic view which was strongly supported by Einstein. According to this view, a scientific theory is a conceptual construction created to capture elements of the reality, which is supposed to exist independently of the observer.  

\n
It is clear that if one accepts the strumentalist point of view then an  explanation {\em for all practical purposes } can be considered  satisfactory  while, on the other hand, a realistic view would force to a deeper analysis of the theory or, eventually, of its interpretation.

\n
For a rather long  period in the scientific community it was spread out the opinion that QM  constrains to reject  the realistic view.  This idea  was based on a consideration of the mathematical formalism and of its interpretation (according to the standard formulation) as an inextricable whole, obscuring the necessary distinction between the two levels.

\n
It was later realized that such  distinction is crucial and that  many different and coherent  interpretations of the mathematical formalism are possible. In particular,  some of these interpretations are compatible with a realistic view.   It should be remarked that the choice of the appropriate interpretation cannot be decided on  a purely scientific level. 
In fact  genuinely philosophical questions are involved in this choice,  and one can only discuss positive or negative aspects of each position without pretending to close the debate.

\n

\n
We want to conclude noticing  that the discussion  on  such foundational  questions can be of interest  not  only at epistemological level but also because it is  a source  of new and possibly stimulating  approaches to scientific problems. At least from this pragmatic point of view  it is relevant  for the concrete development of the scientific knowledge.

\vs
\vs
\vs
\vs
\vs

\vs
\vs
\vs

\end{document}